\documentclass[a4paper,cits]{PoS}

\usepackage{amsmath,amssymb,amsfonts} 
\usepackage[normalem]{ulem}
\usepackage[utf8]{inputenc}
\usepackage[T1]{fontenc}
\usepackage{cancel}
\usepackage{graphicx}
\usepackage{latexsym}
\usepackage{amsmath}
\usepackage{amssymb}
\usepackage{bbm}
\usepackage{epsfig}
\usepackage{epstopdf}
\usepackage{subcaption}
\usepackage{color}
\usepackage{comment}
\usepackage{slashed}
\usepackage{placeins}
\usepackage{cleveref}
\usepackage{setspace}
\usepackage[super]{nth}
\usepackage[dvipsnames]{xcolor}
\usepackage{tikz,wrapfig}
\usetikzlibrary{fadings,calc,decorations.markings}

\newcommand{\dd}{\rm d}

\newcommand{\Nc}{N_{\rm c}}

\newcommand{\Tc}{T_{\rm c}}

\newcommand{\GE}{G_\mathrm{E}}
\newcommand{\tauT}{\tau T}
\newcommand{\GElat}{G_\mathrm{E,\,lat}}
\newcommand{\GEpert}{G_\mathrm{E,\,pert}}
\newcommand{\GEnorm}{G_\mathrm{E,\,norm}}
\newcommand{\GEcont}{G_\mathrm{E,\,cont}}

\newcommand{\ReTr}{\textrm{Re}\,\textrm{Tr}\,}

\newcommand{\be}{\begin{equation}} 
\newcommand{\ee}{\end{equation}}
\newcommand{\bea}{\begin{eqnarray}} 
\newcommand{\eea}{\end{eqnarray}}

\title{Heavy quark momentum diffusion coefficient from the lattice}
\ShortTitle{Heavy quark momentum diffusion coefficient from the lattice}

\author{Nora Brambilla$^a$, \speaker{Viljami Leino}$^a$, Peter Petreczky$^b$, and Antonio Vairo$^a$\\
        $^a$ Technische Universit\"at M\"unchen, Physik Department, James-Franck-Str. 1, 85748 Garching, Germany \\
        $^b$ Physics Department, Brookhaven National Laboratory, Upton, NY 11973, USA \\
 	E-mail: \email{nora.brambilla@ph.tum.de},
        \email{viljami.leino@tum.de},
        \email{petreczk@quark.phy.bnl.gov},
        \email{antonio.vairo@ph.tum.de}}
\author{(TUMQCD collaboration)}

\abstract{We report progress towards computing the heavy quark momentum diffusion coefficient from the correlator
  of two chromoelectric fields attached to a Polyakov loop in pure SU(3) gauge theory.
  Using a multilevel algorithm and tree-level improvement, we study the behavior of the diffusion coefficient as a function
  of temperature in the wide range $1.1<T/T_c<10^4$.}

\FullConference{37th International Symposium on Lattice Field Theory - Lattice2019\\
		16-22 June 2019\\
		Wuhan, China}

\begin{document}

\section{Introduction}
The quark gluon plasma (QGP) produced in heavy ion collisions can be described as a nearly ideal fluid (for a recent review, see Ref.~\cite{Busza:2018rrf}).
In the QGP, the relaxation time of a heavy quark of mass $M$ is expected to be $\sim (M/T) t_{\rm rel}^{\rm light}$, 
with $t_{\rm rel}^{\rm light}$ being the relaxation time of the bulk (light) degrees of freedom in the QGP, and $T$ being the temperature. 
Since the relaxation time of the heavy quark is much larger than the relaxation time of the bulk degrees of freedom,
the heavy quark dynamics can be described by a Langevin equation~\cite{Moore:2004tg}.
In the Langevin equation, the interaction of the heavy quark with the medium is parameterized by 
the drag coefficient $\eta$ and the  heavy quark momentum diffusion coefficient $\kappa$.
They are related by the Einstein equation: $\eta=\kappa/(2MT)$.
Moreover, the heavy quark momentum diffusion coefficient is also a crucial parameter entering the 
evolution equations describing the out-of-equilibrium dynamics of heavy quarkonium in a strongly coupled  QGP~\cite{Brambilla:2016wgg,Brambilla:2017zei}.

The heavy quark momentum diffusion coefficient has been calculated in perturbation theory 
up to next-to-leading order (NLO)~\cite{Moore:2004tg,Svetitsky:1987gq,CaronHuot:2008uh}.
The NLO correction is, however, large, thus questioning the convergence of the perturbative expansion.
Hence, non-perturbative lattice QCD calculations seem to be better suited to determine this quantity.

The direct calculation of the heavy quark momentum diffusion coefficient from lattice QCD is known to be difficult
due to the extremely narrow transport peak, whose width one would need to determine~\cite{Petreczky:2005nh}.
This difficulty can be circumvented by integrating out the heavy quark fields, and relating the heavy quark momentum 
diffusion coefficient to the correlator of two chromoelectric fields~\cite{CaronHuot:2009uh}.
The corresponding spectral function does not have a transport peak, and its small frequency, $\omega$, behaviour 
is smoothly connected to the ultraviolet one~\cite{CaronHuot:2009uh}, which is known at NLO~\cite{Burnier:2010rp}.
Therefore, using the NLO result, one can constrain the functional form of the spectral function used in the analysis
of the lattice correlator in the high $\omega$ region,
while the heavy quark momentum diffusion coefficient is given by the $\omega\rightarrow 0$ limit of the spectral function.
Lattice calculations of $\kappa$ along these lines have been carried out in the deconfined phase of the pure SU(3) 
gauge theory~\cite{Meyer:2010tt,Francis:2011gc,Banerjee:2011ra,Francis:2015daa}.

At sufficiently high temperatures, perturbation theory should describe the heavy quark momentum diffusion coefficient adequately.
Perturbation theory suggests that $\kappa/T^3$ decreases from large values at temperatures close to
the transition temperature $\Tc$ to smaller values as the temperature is increased.
This temperature dependence of $\kappa/T^3$ is phenomenologically important, since a constant value of $\kappa/T^3$
fails to explain simultaneously the elliptic flow parameter $v_2$ for heavy quarks and the nuclear modification factor~\cite{Rapp:2018qla}. 
Previous lattice studies have either considered only a single value of the temperature~\cite{Francis:2015daa},
or focused on a narrow temperature region~\cite{Banerjee:2011ra}.
No significant temperature dependence of $\kappa/T^3$ was found.

\section{Chromoelectric correlator}
For a heavy quark of mass $M\gg\pi T$, the heavy quark effective theory allows to relate the heavy quark momentum diffusion coefficient 
to a purely gluonic observable, the chromoelectric correlator~\cite{CasalderreySolana:2006rq,CaronHuot:2009uh}:
\be\label{eq:gelat}
\GE(\tau) = -\sum_{i=1}^{3} 
 \frac{\left\langle \ReTr\left[U(\beta,\tau)E_i(\tau,\mathbf{0})U(\tau,0)E_i(0,\mathbf{0})\right]\right\rangle}{3\left\langle\ReTr U(\beta,0)\right\rangle}\,,
\ee
where $\beta=1/T$ and $U(\tau_1,\tau_2)$ is the temporal Wilson line connecting $\tau_1$ with $\tau_2$.
The chromoelectric field, in which the coupling has been absorbed $E_i\equiv gE_i$, is discretized on the lattice as~\cite{CaronHuot:2009uh}:
\be\label{eq:elfield}
E_i(\mathbf{x},\tau) = U_i(\mathbf{x},\tau) U_4(\mathbf{x}+\hat{i},\tau) - U_4(\mathbf{x},\tau)U_i(\mathbf{x}+\hat{4})\,.
\ee
In the continuum, the leading order (LO) behavior of $\GE$ is analytically known~\cite{CaronHuot:2009uh}:  
\be\label{eq:gepert}
\frac{\GEpert(\tau)}{g^2C_F} = \pi^2 T^4 \left[\frac{\cos^2(\pi\tau T)}{\sin^4(\pi \tau T)}+\frac{1}{3\sin^2(\pi\tau T)}\right]\,,
\ee
where $C_F=(N_c^2-1)/(2N_c)$ and $N_c=3$ is the number of colors, 
while the NLO result follows from numerically integrating over the NLO spectral function~\cite{Burnier:2010rp}.
By matching the LO continuum expression~\eqref{eq:gepert} with the LO expression in lattice perturbation theory,
one can define a tree-level improved distance $\GEpert(\overline{\tau}) = \GElat^\mathrm{LO}(\tau)$~\cite{Francis:2011gc}. 
We use this tree-level improvement throughout the text usually without any further indication.
Furthermore, the lattice chromoelectric field correlator has to be multiplicatively renormalized $\GEcont(\tau) = Z_\mathrm{E} \GElat(\tau)$. 
For the discretization~\eqref{eq:elfield} the renormalization coefficient $Z_\mathrm{E}$ is known up to NLO~\cite{Christensen:2016wdo}:
$Z_\mathrm{E} = 1+0.13771856909427574(1)g_0^2$.

We compute the discretized chromoelectric correlator~\eqref{eq:gelat} on the lattice using the multilevel algorithm~\cite{Luscher:2001up} with the Wilson gauge action.
To perform the simulations we use the program of Ref.~\cite{Banerjee:2011ra}.
We divide the lattice into 4 sublattices and perform 2000 multilevel updates on each sublattice to compute a single configuration.

\begin{figure}[ht]
  \includegraphics[width=0.49\textwidth]{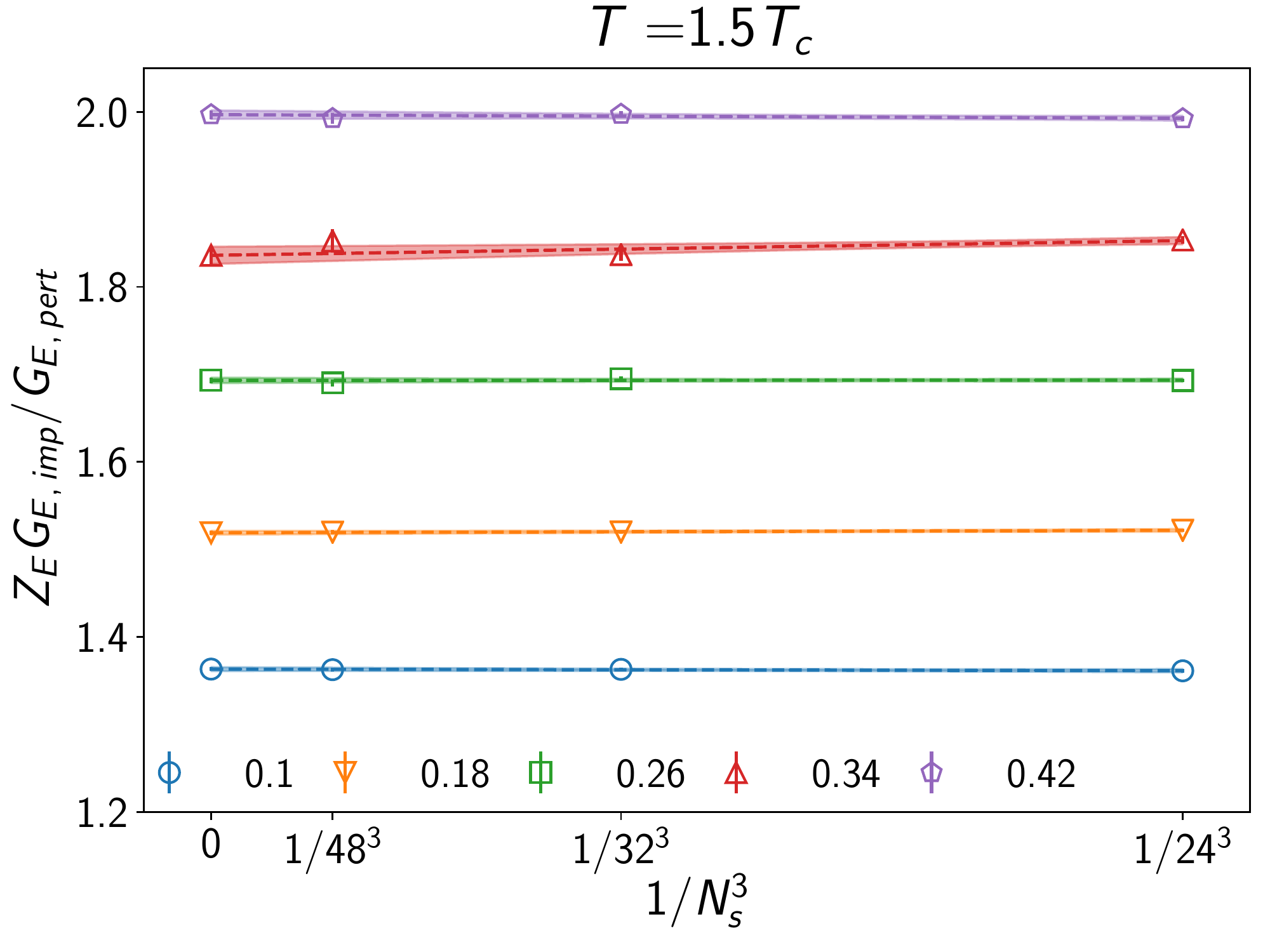}
  \includegraphics[width=0.49\textwidth]{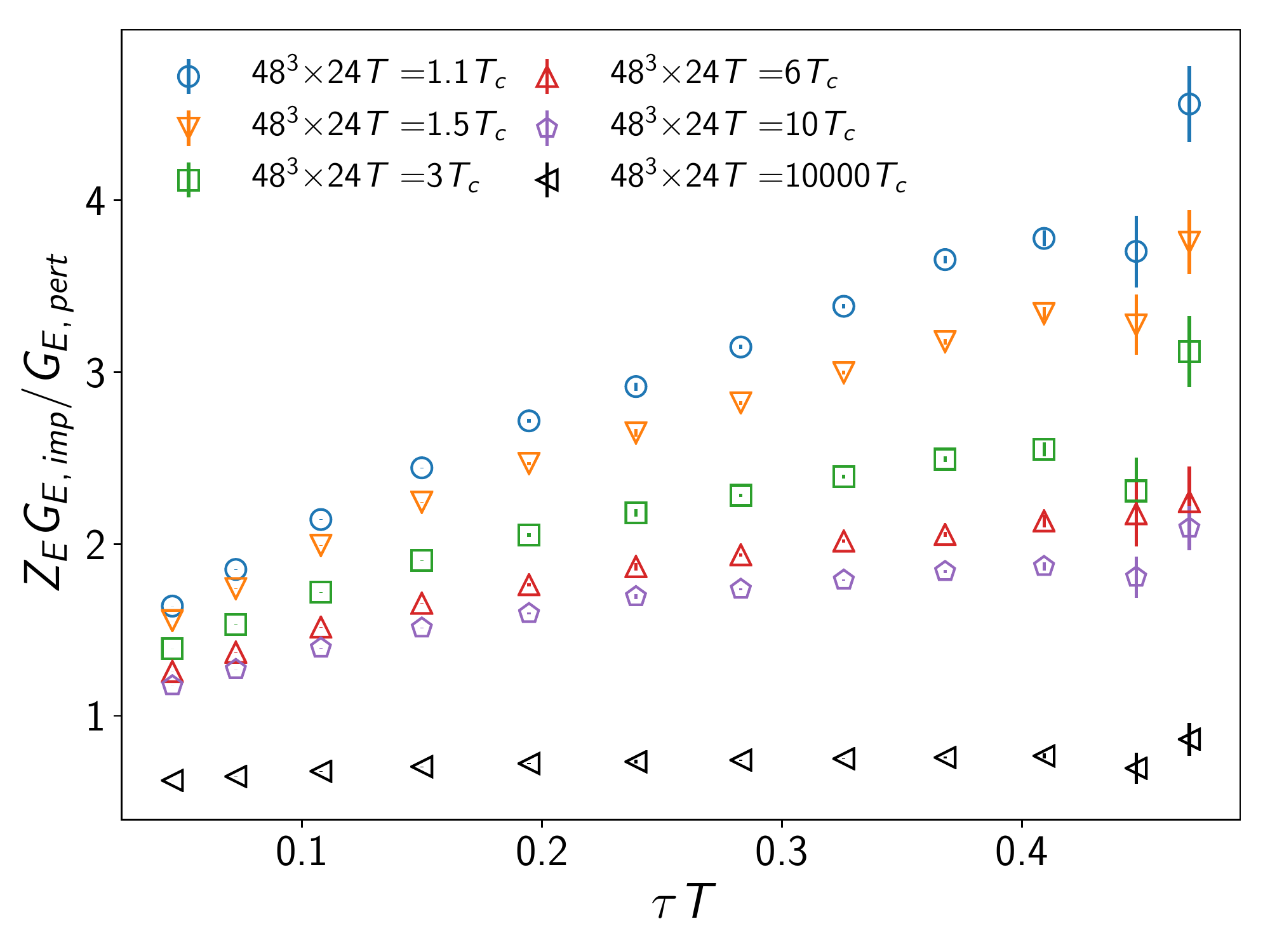}
  \caption[b]{ Left: Discretization effects for a range of $\tauT$ at $T=1.5\Tc$.
               Right: The discretized Euclidean chromoelectric correlator~\eqref{eq:gelat} computed at the largest available volume 
	              for all considered temperatures. $G_{E,\,{\rm imp}}$ is the tree-level improved correlator, see text.}
  \label{fig:disceff} 
\end{figure}

First, we show that finite volume effects are under control by presenting on the left hand side of Fig.~\ref{fig:disceff}
the infinite volume extrapolation done with spatial sizes $N^3_\mathrm{s}=24^3$, $32^3$, $48^3$ and temporal size $N_\mathrm{t}=12$ at $T=1.5\Tc$.
We observe that the extrapolated value differs from the largest volume determination by less than the statistical errors
and, therefore, we will use only $N_\mathrm{s}=48$ for the rest of the analysis and rely on its statistical errors.

\begin{table}[ht]\footnotesize
    \centering
	\setlength\tabcolsep{5pt}
    \begin{tabular}{c|ccccc|ccccc|ccc}
        $T/T_c$ & $N_t\times N_s^3$ & $\beta$ & $N_\mathrm{conf}$ & & $T/T_c$ & $N_t\times N_s^3$ & $\beta$ & $N_\mathrm{conf}$ & & $T/T_c$ & $N_t\times N_s^3$ & $\beta$ & $N_\mathrm{conf}$\\
		\cline{1-4}\cline{6-9}\cline{11-14}                                              
		&     $12 \times 48^3$  & 6.407   & 1350   & &      &     $12 \times 48^3$  & 7.193   & 1579    & & &     $12 \times 48^3$  & 8.211   & 1807   \\
		1.1 & $16 \times 48^3$  & 6.621   & 2623   & &      3   & $16 \times 48^3$  & 7.432   & 1553    & & 10 &  $16 \times 48^3$  & 8.458   & 2769   \\
		&     $20 \times 48^3$  & 6.795   & 1575   & &      &     $20 \times 48^3$  & 7.620   & 1401    & & &     $20 \times 48^3$  & 8.651   & 1613   \\
		&     $24 \times 48^3$  & 6.940   & 2355   & &      &     $24 \times 48^3$  & 7.774   & 1300    & & &     $24 \times 48^3$  & 8.808   & 2241   \\
        \cline{1-4}\cline{6-9}  \cline{11-14}                                        
        &     $12 \times 48^3$  & 6.639   & 1801   & &     &     $12 \times 48^3$  & 7.774   & 1587    & & &      $12 \times 48^3$    & 14.194 & 1039   \\
        1.5 & $16 \times 48^3$  & 6.872   & 2778   & &     6   & $16 \times 48^3$  & 8.019   & 1556    & & $10^4$ & $16 \times 48^3$  & 14.443 & 1157   \\
		&     $20 \times 48^3$  & 7.044   & 1622   & &     &     $20 \times 48^3$  & 8.211   & 1258    & & &      $20 \times 48^3$    & 14.635 & 1139   \\
        &     $24 \times 48^3$  & 7.192   & 2316   & &     &     $24 \times 48^3$  & 8.367   & 1067    & & &      $24 \times 48^3$    & 14.792 & 1190   \\
        \cline{1-4}\cline{6-9}
        2.2  & $12 \times 48^3$  & 6.940   & 1535  & & $2\times10^4 $& $12 \times 48^3$  & 14.792 & 1498 & \multicolumn{5}{c}{}      \\
    \end{tabular}
    \caption{Main simulation parameters and statistics}
    \label{tab:confs}\normalsize
\end{table}

\begin{figure}[ht]
  \includegraphics[width=0.49\textwidth]{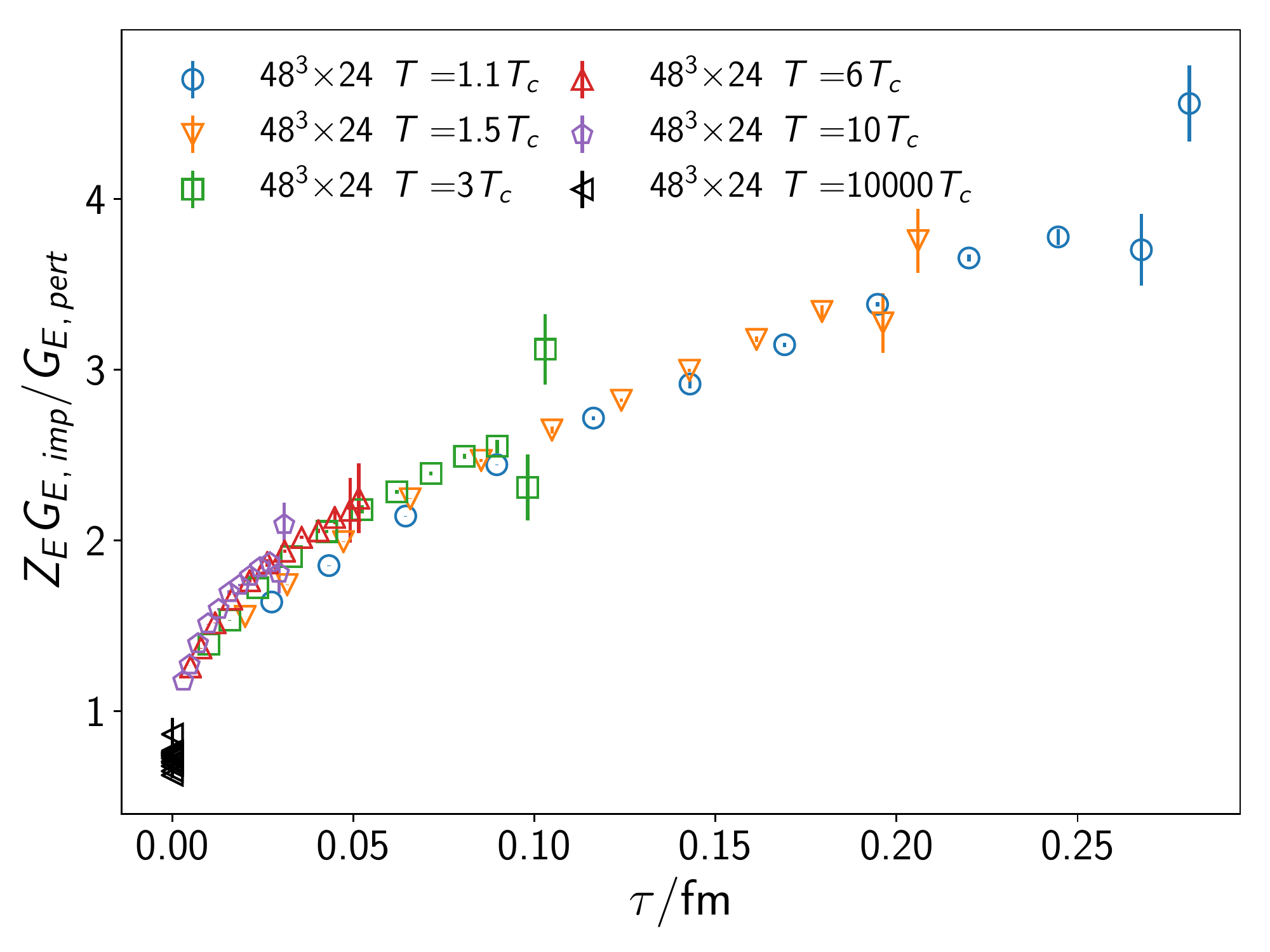}
  \includegraphics[width=0.49\textwidth]{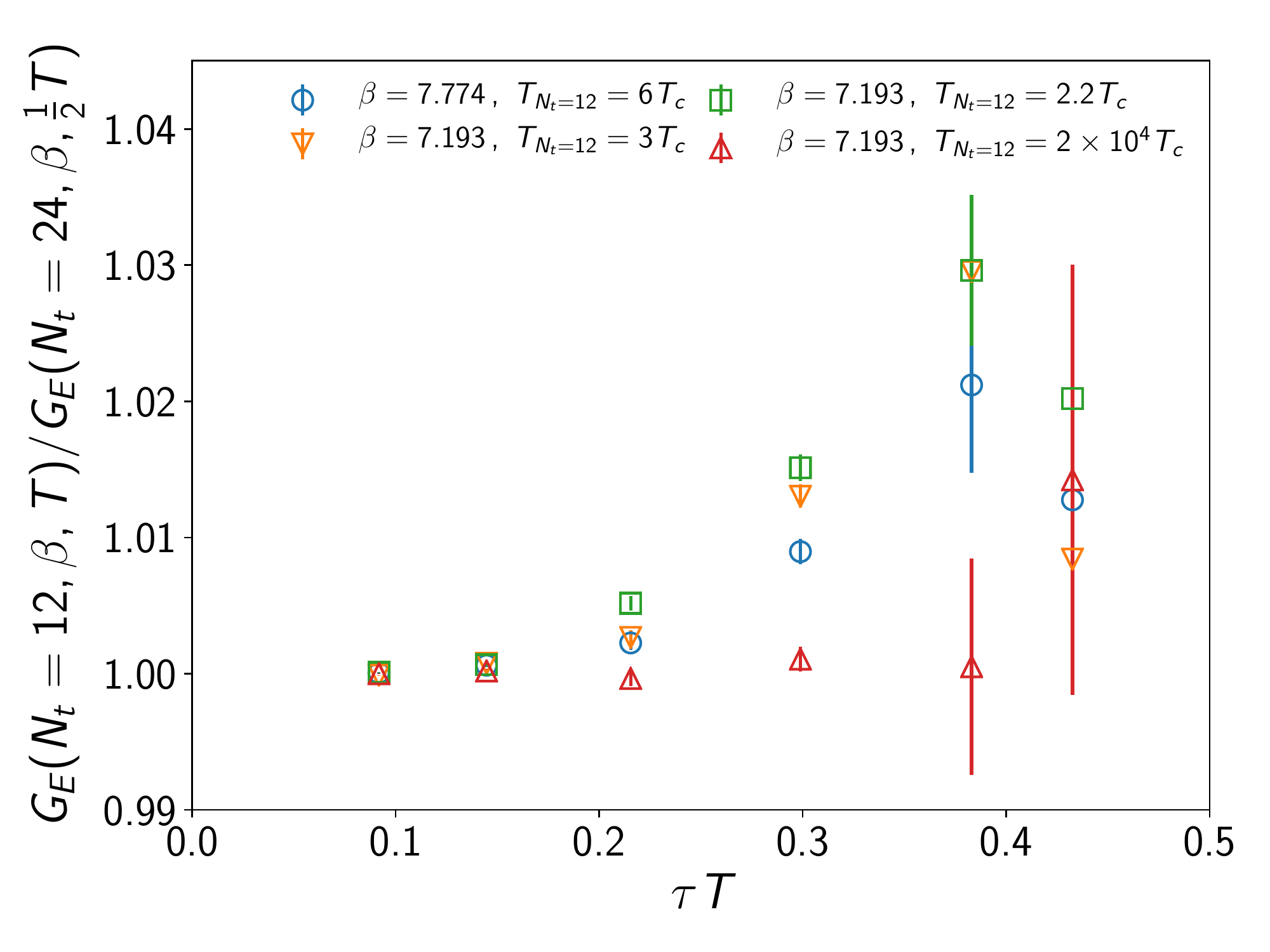}
  \caption[b]{Left: The data of the right hand side of Fig~\ref{fig:disceff} in physical units.
  			  Right: Ratio of simulations with the same $\beta$. 
  }
  \label{fig:fermipert} 
\end{figure}

After the spatial volume has been fixed, we compute the discretized chromoelectric correlator  
on the temporal sizes $N_\mathrm{t}=12$, $16$, $20$, $24$ and temperatures $T=1.1\Tc$, $1.5\Tc$, $3\Tc$, $6\Tc$, $10\Tc$, $10^4\Tc$.
We match the temperatures to the simulation parameter $\beta$ using the method of Ref.~\cite{Francis:2015lha} with scale setting parameter $t_0$, $\Tc \sqrt{t_0}=0.2489(14)$.
The statistics and parameters of our simulations are given in Table~\ref{tab:confs}.
The results are presented in the right hand side of Fig.~\ref{fig:disceff}, 
where the data has been made more readable by normalizing it with~\eqref{eq:gepert}.
Moreover, on the left hand side of Fig.~\ref{fig:fermipert}, we scale the data shown on the right hand side of Fig.~\ref{fig:disceff} to physical units. 
This figure shows that the ratio of the chromoelectric correlator to the free theory result is largely temperature independent
implying that the chromoelectric correlator is dominated by the vacuum part of the spectral function.
To further understand this behavior, we note that when the temperature is doubled then $\beta$ stays the same if the temporal extent is halved.
This allows us to probe the thermal effects by considering the ratio of two simulations with the same lattice spacings.
This is done in the right hand side of Fig.~\ref{fig:fermipert}.
We observe that at small $\tauT$ there is no thermal dependence, while at large $\tauT$ we see more thermal effects as the temperature approaches the transition temperature.
In particular, we see very little temperature dependence in the $T=10^4\Tc$ data.

\begin{figure}[ht]
  \includegraphics[width=0.49\textwidth]{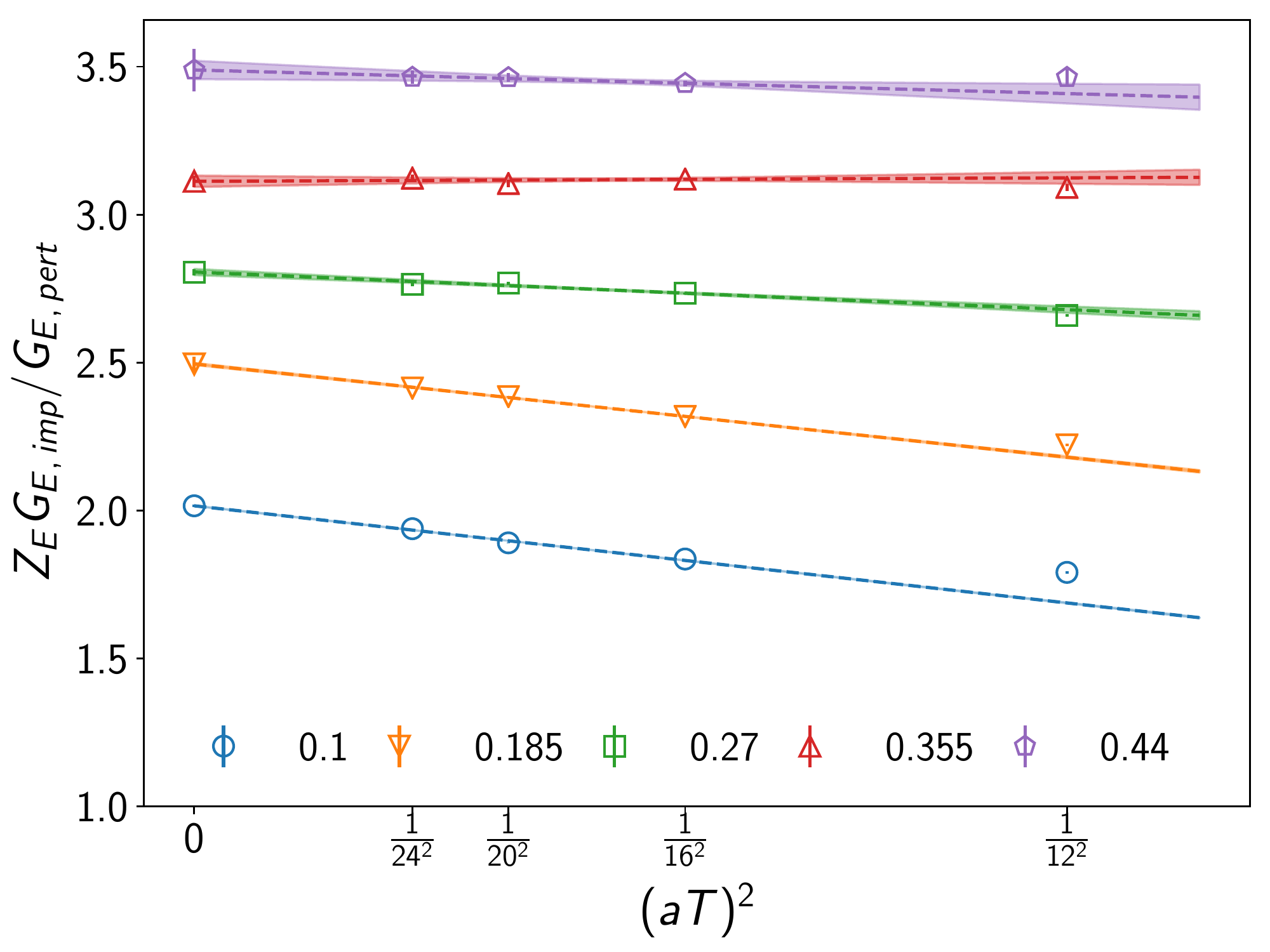}
  \includegraphics[width=0.49\textwidth]{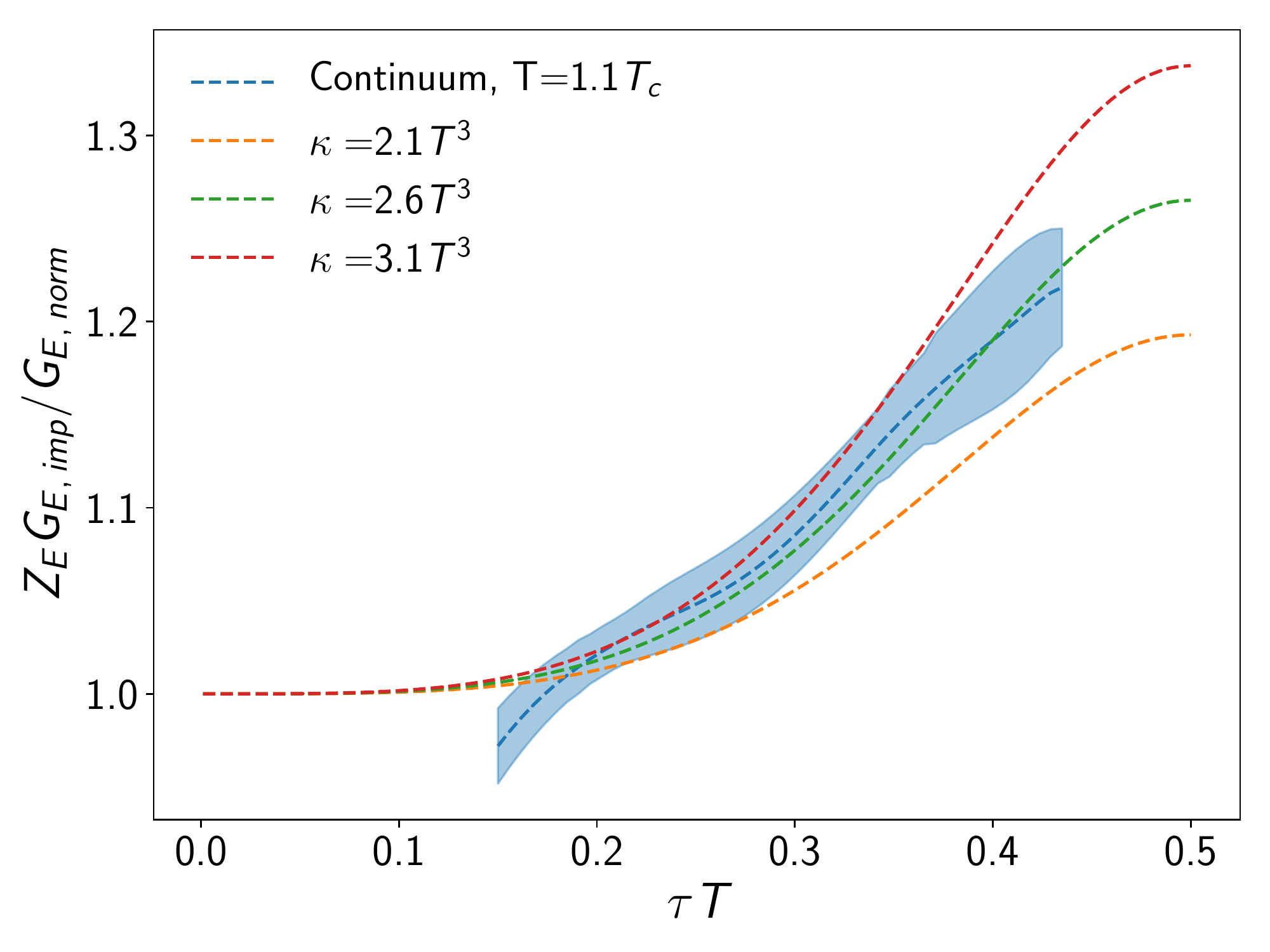}
  \caption[b]{ Left: The continuum limit at selected values of $\tauT$.  
    Right: The continuum limit at $T=1.1\Tc$ compared to the correlator obtained from the ansatz \eqref{eq:anansa} for different values of $\kappa$.
  }
  \label{fig:lim} 
\end{figure}

Next, we perform the continuum limit with the lattices listed in the Table~\ref{tab:confs}.
To do this we interpolate the data over continuous values of $\tauT$ with an \nth{8} order polynomial ansatz.
We perform the continuum limit with the three largest lattice sizes $N_t=16$, $20$, $24$ and check the systematics by also
including the $N_t=12$ lattice to the extrapolation and adding the difference in quadrature.
For $0.18<\tauT<0.45$ we have $\chi^2/\mathrm{d.o.f.}<2.5$ for the larger set of lattices. 
Outside of the $0.18<\tauT<0.45$ range the continuum limit appears less under control.
The continuum limit procedure is illustrated in the left panel of Fig.~\ref{fig:lim} for selected values of $\tauT$.

\section{Heavy quark momentum diffusion coefficient}
In order to compute the heavy quark momentum diffusion coefficient $\kappa$ from the Euclidean correlator \eqref{eq:gelat},
one relates the correlator to the spectral function $\rho(\omega)$:
\be\label{eq:gegegel}
\GE(\tau) = \int_0^\infty \frac{\dd\omega}{\pi}\rho(\omega)\,K(\omega,\tau)\,,\quad
K(\omega,\tau) = \frac{\cosh\left(\frac{\omega}{T}\left(\tauT-\frac{1}{2}\right)\right)}{\sinh\left(\frac{\omega}{2T}\right)}\,,
\ee
which, in turn, is related to $\kappa$~\cite{CaronHuot:2009uh}:
\be\label{eq:kappadeflim}
\kappa = \lim_{\omega\rightarrow0} \frac{2T\rho(\omega)}{\omega}\,.
\ee

The spectral function $\rho(\omega)$ is known up to NLO~\cite{Burnier:2010rp} and for $T=0$ reads:
\begin{align}\label{eq:nlorho}
 \rho_{T=0}^\mathrm{NLO}(\omega)  =  
 \frac{g^2(\mu_\omega) C_F \omega^3}{6\pi} \left[ 1 + 
 \frac{g^2(\mu_\omega)}{(4\pi)^2} \Nc \left( \frac{11}{3} \ln\frac{\mu_\omega^2}{4\omega^2} + \frac{149}{9} - \frac{8\pi^2}{3} \right) \right].
\end{align}
We set the scale $\mu_\omega$ so that for $\omega \gtrsim 0.89T$ the NLO contribution in~\eqref{eq:nlorho} vanishes
and for $\omega\lesssim 0.89T$ the NLO correction to the gauge coupling in EQCD vanishes~\cite{Kajantie:1997tt}, i.e., 
for $\omega \gtrsim  0.89T$ we choose $\displaystyle\ln\mu_\omega = \ln(2\omega) + \frac{24\pi^2-149}{66}$,
and for $\omega \lesssim 0.89T$ we choose $\displaystyle \ln\mu_\omega=\ln(4\pi T) - \gamma_\mathrm{E} -\frac{1}{22}$.
Clearly, this choice is somewhat arbitrary, and in the final analysis one will need to assess the uncertainty related with it.

In order to see how much of the behaviour of the chromoelectric correlator shown in the right panel of Fig.~\ref{fig:disceff}
can be captured by perturbative QCD, in the right panel of Fig.~\ref{fig:lim} we have changed the normalization of the data
from $\GEpert$, which includes only the LO contribution given in Eq.~\eqref{eq:gepert}, 
to $\GEnorm$, which is~\eqref{eq:gegegel} evaluated with the NLO spectral function given in Eq.~\eqref{eq:nlorho}
and with the renormalization scale fixed in the way discussed in the previous paragraph.
With the normalization $\GEnorm$ we expect to see a plateau at low $\tauT$, and, indeed,  we do see an indication of such a behavior. 
However, the plateau is not exactly at 1, indicating possible systematic effects coming from neglected higher order contributions.
To circumvent this we normalize the data to~1~\cite{Meyer:2010tt} at $\tauT=0.18$, 
and add a 1\% systematic error to the normalization from the choice of the normalization point location.

To perform a preliminary extraction of $\kappa$ from our data, we employ for the spectral function the very simple ansatz:
\be\label{eq:anansa}
\rho_\mathrm{ansatz} = \frac{\kappa\omega}{2T}\,\theta(\Lambda-\omega) + \rho_{T=0}^\mathrm{NLO}(\omega)\,\theta(\omega-\Lambda)\,,
\ee
where $\rho_{T=0}^\mathrm{NLO}$  is given by~\eqref{eq:nlorho} and $\Lambda$ is set so that $\kappa\Lambda/(2T) = \rho_{T=0}^\mathrm{NLO}(\Lambda)$.
Computing $\GE$ with the assumption \eqref{eq:anansa} allows to determine the values of $\kappa$ that match the lattice data the best.
This procedure is illustrated on the right hand side of Fig.~\ref{fig:lim}.
To account for the uncertainty related to the choice of the ansatz for the spectral function, we double the errors shown in Fig.~\ref{fig:lim}.
Finally, we find
\bea
2.28 < \frac{\kappa}{T^3} < 3.57  \;&\text{for}&\; T=1.1\Tc\,, \\
1.99 < \frac{\kappa}{T^3} < 2.69  \;&\text{for}&\; T=1.5\Tc\,, \\
1.05 < \frac{\kappa}{T^3} < 2.26  \;&\text{for}&\; T=3\Tc\,, \\
0 < \frac{\kappa}{T^3} < 1.5      \;&\text{for}&\; T=6\Tc\,, \\
0 < \frac{\kappa}{T^3} < 0.91     \;&\text{for}&\; T=10\Tc\,, \\
0 < \frac{\kappa}{T^3} < 0.39     \;&\text{for}&\; T=10^4\Tc\,.
\eea
Our results agree with the existing determinations in the low $T$ regime~\cite{Meyer:2010tt,Francis:2011gc,Banerjee:2011ra,Francis:2015daa,Brambilla:2019tpt}.
Outside of this regime, we observe that $\kappa/T^3$ depends on the temperature, as expected from perturbation theory, becoming smaller at higher ones.

\FloatBarrier
\acknowledgments{
We thank Saumen Datta for sharing with us the multilevel simulation code of Ref.~\cite{Banerjee:2011ra}.
N.B., V.L., and A.V. acknowledge the support from the Bundesministerium f\"ur Bildung und Forschung through project No.~05P2018  
and from the DFG cluster of excellence ORIGINS \href{www.origins-cluster.de}{(www.origins-cluster.de)}.
P.P. has been supported by the U.S. Department of Energy under Contract No.~DE-SC0012704. 
The simulations have been carried out at the computing facilities of the Computational Center for Particle and Astrophysics (C2PAP) in Munich.
}

\bibliographystyle{jhep}
\bibliography{kappa}

\providecommand{\href}[2]{#2}\begingroup\raggedright\begin{thebibliography}{10}

\bibitem{Busza:2018rrf}
W.~Busza, K.~Rajagopal and W.~van~der Schee, \emph{{Heavy Ion Collisions: The
  Big Picture, and the Big Questions}},
  \href{https://doi.org/10.1146/annurev-nucl-101917-020852}{\emph{Ann. Rev.
  Nucl. Part. Sci.} {\bfseries 68} (2018) 339}
  [\href{https://arxiv.org/abs/1802.04801}{{\ttfamily 1802.04801}}].

\bibitem{Moore:2004tg}
G.~D. Moore and D.~Teaney, \emph{{How much do heavy quarks thermalize in a
  heavy ion collision?}},
  \href{https://doi.org/10.1103/PhysRevC.71.064904}{\emph{Phys. Rev.}
  {\bfseries C71} (2005) 064904}
  [\href{https://arxiv.org/abs/hep-ph/0412346}{{\ttfamily hep-ph/0412346}}].

\bibitem{Brambilla:2016wgg}
N.~Brambilla, M.~A. Escobedo, J.~Soto and A.~Vairo, \emph{{Quarkonium
  suppression in heavy-ion collisions: an open quantum system approach}},
  \href{https://doi.org/10.1103/PhysRevD.96.034021}{\emph{Phys. Rev.}
  {\bfseries D96} (2017) 034021}
  [\href{https://arxiv.org/abs/1612.07248}{{\ttfamily 1612.07248}}].

\bibitem{Brambilla:2017zei}
N.~Brambilla, M.~A. Escobedo, J.~Soto and A.~Vairo, \emph{{Heavy quarkonium
  suppression in a fireball}},
  \href{https://doi.org/10.1103/PhysRevD.97.074009}{\emph{Phys. Rev.}
  {\bfseries D97} (2018) 074009}
  [\href{https://arxiv.org/abs/1711.04515}{{\ttfamily 1711.04515}}].

\bibitem{Svetitsky:1987gq}
B.~Svetitsky, \emph{{Diffusion of charmed quarks in the quark-gluon plasma}},
  \href{https://doi.org/10.1103/PhysRevD.37.2484}{\emph{Phys. Rev.} {\bfseries
  D37} (1988) 2484}.

\bibitem{CaronHuot:2008uh}
S.~Caron-Huot and G.~D. Moore, \emph{{Heavy quark diffusion in QCD and N=4 SYM
  at next-to-leading order}},
  \href{https://doi.org/10.1088/1126-6708/2008/02/081}{\emph{JHEP} {\bfseries
  02} (2008) 081} [\href{https://arxiv.org/abs/0801.2173}{{\ttfamily
  0801.2173}}].

\bibitem{Petreczky:2005nh}
P.~Petreczky and D.~Teaney, \emph{{Heavy quark diffusion from the lattice}},
  \href{https://doi.org/10.1103/PhysRevD.73.014508}{\emph{Phys. Rev.}
  {\bfseries D73} (2006) 014508}
  [\href{https://arxiv.org/abs/hep-ph/0507318}{{\ttfamily hep-ph/0507318}}].

\bibitem{CaronHuot:2009uh}
S.~Caron-Huot, M.~Laine and G.~D. Moore, \emph{{A Way to estimate the heavy
  quark thermalization rate from the lattice}},
  \href{https://doi.org/10.1088/1126-6708/2009/04/053}{\emph{JHEP} {\bfseries
  04} (2009) 053} [\href{https://arxiv.org/abs/0901.1195}{{\ttfamily
  0901.1195}}].

\bibitem{Burnier:2010rp}
Y.~Burnier, M.~Laine, J.~Langelage and L.~Mether, \emph{{Colour-electric
  spectral function at next-to-leading order}},
  \href{https://doi.org/10.1007/JHEP08(2010)094}{\emph{JHEP} {\bfseries 08}
  (2010) 094} [\href{https://arxiv.org/abs/1006.0867}{{\ttfamily 1006.0867}}].

\bibitem{Meyer:2010tt}
H.~B. Meyer, \emph{{The errant life of a heavy quark in the quark-gluon
  plasma}}, \href{https://doi.org/10.1088/1367-2630/13/3/035008}{\emph{New J.
  Phys.} {\bfseries 13} (2011) 035008}
  [\href{https://arxiv.org/abs/1012.0234}{{\ttfamily 1012.0234}}].

\bibitem{Francis:2011gc}
A.~Francis, O.~Kaczmarek, M.~Laine and J.~Langelage, \emph{{Towards a
  non-perturbative measurement of the heavy quark momentum diffusion
  coefficient}}, \href{https://doi.org/10.22323/1.139.0202}{\emph{PoS}
  {\bfseries LATTICE2011} (2011) 202}
  [\href{https://arxiv.org/abs/1109.3941}{{\ttfamily 1109.3941}}].

\bibitem{Banerjee:2011ra}
D.~Banerjee, S.~Datta, R.~Gavai and P.~Majumdar, \emph{{Heavy Quark Momentum
  Diffusion Coefficient from Lattice QCD}},
  \href{https://doi.org/10.1103/PhysRevD.85.014510}{\emph{Phys. Rev.}
  {\bfseries D85} (2012) 014510}
  [\href{https://arxiv.org/abs/1109.5738}{{\ttfamily 1109.5738}}].

\bibitem{Francis:2015daa}
A.~Francis, O.~Kaczmarek, M.~Laine, T.~Neuhaus and H.~Ohno,
  \emph{{Nonperturbative estimate of the heavy quark momentum diffusion
  coefficient}}, \href{https://doi.org/10.1103/PhysRevD.92.116003}{\emph{Phys.
  Rev.} {\bfseries D92} (2015) 116003}
  [\href{https://arxiv.org/abs/1508.04543}{{\ttfamily 1508.04543}}].

\bibitem{Rapp:2018qla}
A.~Beraudo et~al., \emph{{Extraction of Heavy-Flavor Transport Coefficients in
  QCD Matter}},
  \href{https://doi.org/10.1016/j.nuclphysa.2018.09.002}{\emph{Nucl. Phys.}
  {\bfseries A979} (2018) 21}
  [\href{https://arxiv.org/abs/1803.03824}{{\ttfamily 1803.03824}}].

\bibitem{CasalderreySolana:2006rq}
J.~Casalderrey-Solana and D.~Teaney, \emph{{Heavy quark diffusion in strongly
  coupled N=4 Yang-Mills}},
  \href{https://doi.org/10.1103/PhysRevD.74.085012}{\emph{Phys. Rev.}
  {\bfseries D74} (2006) 085012}
  [\href{https://arxiv.org/abs/hep-ph/0605199}{{\ttfamily hep-ph/0605199}}].

\bibitem{Christensen:2016wdo}
C.~Christensen and M.~Laine, \emph{{Perturbative renormalization of the
  electric field correlator}},
  \href{https://doi.org/10.1016/j.physletb.2016.02.020}{\emph{Phys. Lett.}
  {\bfseries B755} (2016) 316}
  [\href{https://arxiv.org/abs/1601.01573}{{\ttfamily 1601.01573}}].

\bibitem{Luscher:2001up}
M.~Lüscher and P.~Weisz, \emph{{Locality and exponential error reduction in
  numerical lattice gauge theory}},
  \href{https://doi.org/10.1088/1126-6708/2001/09/010}{\emph{JHEP} {\bfseries
  09} (2001) 010} [\href{https://arxiv.org/abs/hep-lat/0108014}{{\ttfamily
  hep-lat/0108014}}].

\bibitem{Francis:2015lha}
A.~Francis, O.~Kaczmarek, M.~Laine, T.~Neuhaus and H.~Ohno, \emph{{Critical
  point and scale setting in SU(3) plasma: An update}},
  \href{https://doi.org/10.1103/PhysRevD.91.096002}{\emph{Phys. Rev.}
  {\bfseries D91} (2015) 096002}
  [\href{https://arxiv.org/abs/1503.05652}{{\ttfamily 1503.05652}}].

\bibitem{Kajantie:1997tt}
K.~Kajantie, M.~Laine, K.~Rummukainen and M.~E. Shaposhnikov, \emph{{3-D SU(N)
  + adjoint Higgs theory and finite temperature QCD}},
  \href{https://doi.org/10.1016/S0550-3213(97)00425-2}{\emph{Nucl. Phys.}
  {\bfseries B503} (1997) 357}
  [\href{https://arxiv.org/abs/hep-ph/9704416}{{\ttfamily hep-ph/9704416}}].

\bibitem{Brambilla:2019tpt}
N.~Brambilla, M.~A. Escobedo, A.~Vairo and P.~Vander~Griend, \emph{{Transport
  coefficients from in medium quarkonium dynamics}},
  \href{https://doi.org/10.1103/PhysRevD.100.054025}{\emph{Phys. Rev.}
  {\bfseries D100} (2019) 054025}
  [\href{https://arxiv.org/abs/1903.08063}{{\ttfamily 1903.08063}}].

\end{thebibliography}\endgroup

\end{document}